# Transparent conducting silver nanowire networks

*Jorik van de Groep\*, Pierpaolo Spinelli and Albert Polman*

Center for Nanophotonics, FOM Institute AMOLF

Science Park 104, 1098 XG, Amsterdam, The Netherlands

AUTHOR EMAIL ADDRESS: *groep@amolf.nl

ABSTRACT: We present a transparent conducting electrode composed of a periodic two-dimensional network of silver nanowires. Networks of Ag nanowires are made with wire diameters of 45-110 nm and pitch of 500, 700 and 1000 nm. Anomalous optical transmission is observed, with an averaged transmission up to 91% for the best transmitting network and sheet resistances as low as 6.5 Ω/sq for the best conducting network. Our most dilute networks show lower sheet resistance and higher optical transmittance than an 80 nm thick layer of ITO sputtered on glass. By comparing measurements and simulations we identify four distinct physical phenomena that govern the transmission of light through the networks: all related to the excitation of localized surface plasmons and surface plasmon polaritons on the wires. The insights given in this paper provide the key guidelines for designing high-transmittance and low-resistance nanowire electrodes for optoelectronic devices, including thin-film solar cells. For these latter, we discuss the general design principles to use the nanowire electrodes also as a light trapping scheme.

KEYWORDS: Transparent conducting electrode, nanowires, anomalous transmission, sheet resistance, plasmonics, light trapping.

MANUSCRIPT TEXT:

**Introduction**

Transparent conducting electrodes (TCEs) are essential components in many optoelectronic devices, including LEDs, electronic displays and solar cells. Their unique combination of optical transparency and electrical conductivity provides the possibility to extract electrical carriers while transmitting light through the layer. Indium-Tin-Oxides (ITOs) are commonly used in many applications, but suffer from several important drawbacks, such as high materials costs,[1,2] scarcity of indium, damaging of organic substrates during sputtering,[1] and brittleness.[1-4] An important further drawback for applications in solar cells is the strong absorption of ITO in the UV/blue spectral range.[5,6] These problems motivate the search for new TCE materials. Presently, many new materials and geometries are being investigated, including macroscopic metallic grids (down to sizes of tens of microns),[4,7] one-dimensional (nano-)imprinted metal electrodes, [2,3,8] solution-processed random metal nanowire (NW) meshes,[9-13] graphene,[14,15] and aluminum nanomeshes fabricated through self-assembly of silica spheres.[16] Each of these materials geometries represents a trade-off between optical transmittance and electrical conduction. Thicker and denser layers provide better electrical conduction but reduce the optical transmission and vice versa. Furthermore, the compatibility with optoelectronic device fabrication techniques limits the applicability of several of these materials.

Recently, a detailed numerical analysis of the physics of the transmission of light through 1-D and 2-D metal NW networks has shown how a combination of guided optical modes in Metal-Insulator-Metal (MIM) waveguides (formed by the NWs) and surface plasmon polaritons (SPPs) on the NWs determine the optimal design of such networks.[1] Thin, high aspect ratio wires were evaluated in simulations to have optimum transmission without losing electrical conductance. The 1-D design has been realized experimentally,[2] yielding performances comparable to ITO. This structure, however is only effective for one polarization. Two dimensional periodic networks of sub-wavelength nanowires and wavelength-sized pitch have been realized before,[17] however these were optimized for absorption in the metal nanowires. So far, there is no experimental study on 2-D sub-wavelength NW networks with wavelength-sized pitch optimized for high optical transmission and low electrical resistivity.

In this paper, we simulate and fabricate a 2-D network of silver NWs, and study the performance of such NW networks as transparent conducting electrodes. By simulating different wire widths, we recognize four different scattering mechanisms that govern the transmittance of the network, and also observe these experimentally. We show that several additional physical effects to the ones described in ref. [1] come into play when considering 2D NW networks with wire spacing on the scale of the wavelength of light. Measurements show that our best geometries show higher optical transmittance and lower electrical resistance than an 80 nm thick layer of ITO.

**Optical characterization: experiment**

Figure 1a shows a sketch of a typical NW network design used in this work. The total size of the network was 50×50 μm, and electrical contact pads with a size of 80×50 μm were designed on two opposing sides of the network to allow electrical characterization. The Ag NW networks were fabricated using electron beam lithography. A SiO$_2$ substrate was used to allow both high optical transmission and electrical insulation. Networks of NWs with 500, 700 and 1000 nm wire spacing ($\Lambda$) were written with a 30 kV electron beam (7.5 μm aperture, ~0.018 nA current, ~2.7 μs dwell time, ~4.8 nm step size) in positive resist (ZEP) using single pixel line exposure. The dose was varied between 400 and 1000 pC/cm to obtain wires with different widths. After development, ~3 nm of germanium (at a rate of 0.2 Å/s) and 50±5 nm of silver (0.5 Å/s) were deposited using thermal evaporation. The layer thickness defines the height of the wires in the network. The thin layer of germanium underneath the silver reduces the grain size of the silver and thereby gives rise to high quality silver NWs.[18-20] Lift-off is performed by soaking in anisole (methoxybenzene) for 24 hours, followed by 1:30 minutes of ultra-sonification in anisol.

Figure 1b,c shows Scanning Electron Microscopy (SEM) images of the $\Lambda$ = 500 nm sample with 55 nm wire width. The image shows the excellent uniformity of the network over a large area. The inset in Fig. 1b shows the high quality of the individual NWs. The NW width for different networks made using different e-beam exposures ranged from 45-110 nm. Figure 1c shows the wire network is well connected to the electrical contact pads. Figure 1d shows an Atomic Force Microscopy (AFM) image of the network in Fig. 1b,c. The inset shows a line scan from which the NW height is determined to be 60 ± 5 nm.

Optical transmission measurements were performed using an integrating sphere setup. A supercontinuum white light source (Fianium) was used to illuminate the networks close to normal incidence. A microscope setup with a 20× objective (NA = 0.4) was used to focus the unpolarized light into a 10 μm spot in the center of the NW networks. A large core diameter detection fibre connected the integrating sphere with a spectrometer consisting of a spectrograph and a Si CCD array. Measurements were performed in the spectral range from 420 nm $< \lambda <$ 970 nm.

Figure 2 shows the results of the optical transmission measurements. In Fig. 2a the normalized transmission (defined as the transmission through the network normalized to the transmission through a bare glass reference sample) is shown as a function of wavelength for a $\Lambda$ = 500 nm network with different wire width in the range $w$ = 45-110 nm (colors). Several clear features can be observed. First, an overall reduction in transmittance is observed for larger $w$, which is attributed to the increase in surface area that is covered by silver as $w$ increases. Second, a deep and spectrally broad dip in transmittance is observed around $\lambda$ = 450-550 nm which shifts to larger wavelengths for larger $w$. This feature is attributed to the excitation of the localized surface plasmon resonance (LSPR) of individual Ag NWs. Third, a sharp dip in transmittance is observed for 750 $< \lambda <$ 850 nm that shifts to shorter wavelength as $w$ increases. This spectrally sharp feature is attributed to the excitation of propagating surface plasmon polaritons (SPPs) along the individual wires. Finally, a downward slope is observed in the transmittance for $\lambda > 850$ nm that decreases for larger wire diameters. This decrease in transmission is attributed to the cut-off of the fundamental $TE_0$ waveguide mode that is supported by the metal-insulator-metal (MIM) waveguides formed by the air gaps in between the wires.[1] Each of these features will be discussed in more detail in the simulations section.

SEM images were used to determine the fraction of the surface area that is covered by the silver network for all wire diameters and each pitch. We define the clear aperture of a network as the fraction of surface are that is not covered by silver. Based on simple geometrical optics, one would expect the transmittance of the network to be equal to the clear aperture. The inset of Fig. 2a shows the same transmission curves as in Fig. 2a, normalized to the clear aperture of each network. Anomalous transmission (i.e., transmission larger than the clear aperture) is observed for all wire diameters at wavelengths above the LSPR wavelength. The average transmission, weighted for the AM1.5 solar spectrum and normalized to the clear aperture, is found to be ~1.1 for all wire diameters. The anomalous transmission is a result of the fact that the individual wires of the network are sub-wavelength structures that strongly scatter light.[21-23] Unlike bulk metal structures, sub-wavelength structures can have large scattering cross sections compared to their geometrical cross section, due to their plasmonic behavior. The fraction of forward-scattered light depends on the local density of states (LDOS) of the top and bottom medium, as stated by Fermi's Golden Rule. As the substrate has a higher refractive than air, and thus a higher LDOS, the nanowires will preferentially scatter into the substrate.[24] The asymmetric shape of the transmittance in the inset of Fig. 2a can be understood from the Fano interference between light scattered by the LSPR of the individual wires and the directly transmitted light.[25-28] Fano interference occurs when light scattered from a spectrally narrow resonance (here the LSPR) interferes with a broad continuum of states (the broadband directly transmitted light), as a result of the different phase pick-up at scattering. Light scattered at wavelengths below the LSPR wavelength interferes destructively with the directly transmitted light, giving rise to a dip in transmission, whereas at wavelengths above the LSPR resonance the scattered light interferes constructively with the directly transmitted light, thus resulting into an increase in transmission.

To compare the optical transmittance through the Ag NW networks to that of ITO we used an 80 nm thick layer of ITO that was sputtered on glass. This is a standard thickness used on amorphous Si and organic solar cells.[6,29] In Fig. 2b we compare the transmittance of the highest transmitting network ($w$ = 45 nm) with three different pitches ($\Lambda$ = 500 (blue), 700 (purple) and 1000 nm (pink)) with the normalized transmission of ITO (gray). The NW networks show better transmittance than ITO over a broad spectral range, except near the LSPR located around $\lambda$ = 400 nm. To compare the transmittance of the Ag NW networks and ITO for photovoltaic applications we calculate the average transmission weighted for the AM1.5 solar spectrum. The part of the spectrum considered in this measurement is highlighted in the plotted spectrum that is shown as an inset in Fig. 2b. As can be seen, this spectral region accounts for a large fraction of the AM1.5 spectral intensity. The average transmission values are listed in the inset in Fig. 2b. The 500 nm pitched network has a transmittance that is equal to that of ITO (88%). The networks with 700 and 1000 nm pitch show slightly higher transmittance: 90% and 91% respectively. This shows that the NW networks have an optical transmittance equal or better than ITO, depending on the NW spacing.

**Optical characterization: simulations**

Finite-difference time-domain (FDTD) simulations[30] were used to further investigate the physical phenomena responsible for the transmission of the NW networks. In the simulations Ag wires with width $w$, height $h$ and pitch $\Lambda$ = 500 nm were positioned on a semi-infinite $SiO_2$ substrate. In the simulations a broad-band ($\lambda$ = 400-1100 nm) plane wave at normal incidence is used as a source, polarized along one of the wire orientations. Power monitors, positioned 1 nm below the air-$SiO_2$ interface and 500 nm above the silver network, are used to determine the transmission and reflection at each wavelength. Perfectly matched layer (PML) boundary conditions were used in the vertical direction to prevent unphysical scattering at the edge of the simulation box. Periodic boundary conditions were used in both in-plane dimensions to simulate an infinite network. Optical constants for silver were obtained from a combined Drude, Lorentz, and Debye model fitted to the data from Palik[31]; optical constants for $SiO_2$ were also taken from Palik.

Figure 3a shows the simulated transmittance $T$ (solid) and reflectance $R$ (dashed) as a function of wavelength for a network consisting of wires with $h$ = 50 nm and a width ranging from $w$ = 30-100 nm (different colors). The wavelength equal to the wire pitch (500 nm), is indicated by the dashed vertical line in gray. A high transmittance is observed with several sharp and broad dips, governed by four different scattering mechanisms.

The broad dip in transmission in the 400-500 nm spectral range shows a red-shift for increasing wire width, as was also observed in the experiment (Fig. 2a). This dip is attributed to the excitation of the LSPR due to oscillations of free electrons in the transverse direction to the individual NWs (see sketch in inset of Fig. 3c). To prove this, we simulate the transmission of an infinitely long single wire in air by introducing PML boundary conditions in the in-plane direction normal to the wire axis. Figure 3c shows the transmission of a single wire with $w$ = 80 nm and $h$ = 50 nm, corresponding to the wire geometry for the blue curve in Fig. 3a. Both curves show a similar dip in transmittance, and a red-shift of the dip position is observed for the wire network (Fig. 3a) compared to a single wire (Fig. 3c) due to inter-wire coupling and the presence of the substrate. Further confirmation of the LSPR origin of the dip is the red-shift of the resonant wavelength for larger $w$ observed in Fig. 3a, which is caused by the reduced restoring force on the oscillating electrons for larger wire width. This dip was also observed in ref. [1], where it is attributed to the excitation of two coupled SPP modes that propagate on the top and bottom interface of the wires. However, our analysis shows that instead it is a result of the excitation of the LSPR. This is further confirmed by the fact that it also appears in the transmission of a single wire (Fig.

3c) onto which SPPs can not be excited by an incident plane wave because of in-plane momentum mismatch.

The second noticeable feature in the transmission spectra in Fig. 3a is a step-like increase in transmission for all wire widths that occurs for wavelengths just above the wire spacing ($\Lambda$ = 500 nm). This is a result of the excitation of ±1 diffraction orders that are directed into the substrate for $\lambda > \Lambda$. At $\lambda$ = 500 nm, the ±1 diffraction orders are scattered parallel to the surface, giving rise to the strong collective excitation of LSPR modes (Rayleigh anomaly).[32] This increased interaction between the incoming light and the NWs gives rise to strong absorption of light in the NWs, as confirmed by NW absorption spectra shown for different wire widths in Fig. 3b. Strong absorption is observed for all widths at $\lambda$ = 500 nm as a result of the collective excitation of the LSPR, with the strongest absorption observed for the largest width. The interference of the broad LSPR mode and the spectrally sharp Rayleigh anomaly (RA) gives rise to a Fano-lineshape in the transmission around $\lambda$ = 500 nm. In the experiment, this feature is small and only observable for the two largest wire widths. This is a direct result of the fact that the sample was illuminated through a microscope objective and the incident light is thus not perfectly collimated. This broadens the RA over a large spectral range.

A third characteristic feature in Fig. 3a is the gradually decreasing transmission for wavelengths above ~750 nm; an effect that becomes stronger for larger $w$. As also described in ref. [1] for one-dimensional networks, this is due to the cut-off of the fundamental $TE_0$ mode supported by the vertically oriented MIM waveguide formed by the wires and the air in between. The decrease in transmission is relatively weak for two reasons. First, in the square network geometry light can couple to both $TE_0$ and $TM_0$ modes. Indeed, a modal dispersion calculation for an infinite MIM waveguide with similar insulator thickness shows a well pronounced cut-off for the $TE_0$ mode, but not for the $TM_0$ mode (data not shown here). Second, the cut-off is not well pronounced due to the very limited waveguide length in the normal direction, which is equal to the wire height (50 nm). The decrease in transmittance is also observed in experiments for wavelengths above 800 nm (see Fig. 2a).

Finally, a sharp dip in transmission is observed around $\lambda$ = 800 nm that blue-shifts and is more pronounced for larger wire width. In contrast to the dips in the blue spectral range, this sharp dip is not accompanied by a large peak in reflectance. This indicates strong absorption takes place in the silver NWs, as is confirmed by the spectra in Fig. 3b. This dip in transmission is caused by the excitation of SPPs propagating in the longitudinal direction along the wires. The lossy nature of these SPP modes causes the peak in absorption. The two-dimensional periodic nature of our structure provides the in-plane momentum required to excite the SPP mode. The wires oriented perpendicular to the electric field component of the incoming light act as a grating to provide the in-plane momentum required to couple to a SPP propagating in the direction parallel to the electric field component (see sketch in inset of Fig 3b). Larger wire widths lead to an increase in the scattering cross section of the wires such that light can couple more efficiently to the SPP mode, which explains why the dip is more pronounced for larger $w$. Furthermore, the SPP mode is more confined in the in-plane direction for wires with smaller $w$, resulting in a higher effective mode index. This explains the red-shift of the SPP transmission dip for thinner wires in Fig. 3a.

To further corroborate the SPP excitation mechanism in the Ag NW networks, we use momentum matching calculations to relate the spectral location of the dip to the dispersion curve of the SPP mode. Electromagnetic Boundary Element Method (BEM) calculations were performed using the BEM2D program[33,34] to calculate the dispersion curves for the confined SPP modes on a 50 nm and 100 nm wide infinitely long wire on SiO$_2$. The dispersion curve was determined by finding local maxima in the local density of optical states (LDOS) 10 nm above the wire, for a broad range of $k_0$ and $\beta$. The curves are shown in Fig. 3d together with light line in air. An increase in mode index is observed for narrower wires, as expected. Coupling to these modes through the periodic NW grating requires momentum matching according to $2\pi m / \Lambda = \beta$, where $m = 1, 2, \ldots$. The dashed lines in Fig. 3d show that coupling to the SPP can occur at $\lambda$ = 836 nm for *w* = 50 nm and $\lambda$ = 793 nm for *w* = 100nm. These wavelengths are in good agreement with the wavelengths of the transmission dips in Fig. 3a ($\lambda$ = 849 nm and 781 nm respectively).

Additional proof for the coupling to SPP modes can be obtained from FDTD by monitoring the near field intensity 5 nm above the surface in the middle of wires perpendicular (y-parallel) and parallel (x-parallel) to the electric field orientation. As explained above, the wires perpendicular to the electric field orientation act as a grating that allows coupling to a SPP parallel to the electric field orientation, such that the near field intensity should show a peak at $\lambda$ = 781 nm for the x-parallel wires only. The near field intensity as a function of wavelength is shown as an inset in Fig. 1d, where the blue line corresponds to the y-parallel and the orange line to the x-parallel. Indeed, a sharp peak at $\lambda$= 781 nm is observed for the orange line only, confirming that the dip is a result of coupling to the SPP mode (indicated by arrow).

The dip in transmission due to coupling to the NW SPP mode and the blue-shift for thicker wires are also clearly observed in the experiment (Fig. 2a). However, in the experiment the dips are spectrally broadened, which we attribute to small variations in wire width and angle of incidence of the light. BEM calculations using the measured wire width and height are used to find the spectral location of the dip as predicted by momentum matching calculations. The results are plotted as the gray dashed line in Fig. 2a. The measured data shows good qualitative agreement with the calculation, although the data is slightly red-shifted. We attribute this to small differences in the wire geometry (cross-sectional shape) and optical constants. Finally, the analysis presented here also explains why a dip due to the coupling to the NW SPP mode was not observed for the larger pitches in Fig. 2b. For these networks the momentum matching condition occurs at longer wavelengths outside the spectral range detected in experiments.

**Electrical characterization: experiment**

The electrical resistance of the Ag NW networks was measured with a four probe experiment such that the probe contact resistance is eliminated. Two probes were positioned on each contact pad and I-V - curves were measured by ramping the current through the network from -1 to 1 mA and measuring the voltage over the network. Typical I-V measurements are shown in the inset of Fig. 4. The graph shows measured data (dots) for networks with a $w$ = 45 nm and $\Lambda$ = 500 (blue), 700 (green) and 1000 nm (orange). Linear fits of the data reflect the Ohmic conduction through the network. The fitted resistances are 17.2 Ω/sq, 27.0 Ω/sq and 38.7 Ω/sq for the $\Lambda$ = 500, 700 and 1000 nm network respectively. According to Kirchoff's rules, the sheet resistance $R_s$ for a square wire network with $N \times N$ wires is given by

$$R_s = \frac{N}{N+1} R_{wire} = \frac{N}{N+1} \frac{\rho L}{wh},$$
(1)

where $\rho$ is the resistivity of silver, $L$ is the length of the wire (equal to network pitch), $w$ is the wire width and $h$ is the height of the wire. For network dimensions used in the experiment ($N$ = 50 to 100) the first term is close to unity, such that $R_s = \rho L / wh$. Indeed, the measured sheet resistances scale linearly with network pitch ($L$).

Figure 4 shows the measured sheet resistance (dots) as a function of wire width for networks with $\Lambda$ = 500 (blue), 700 (green) and 1000 nm (orange). The dashed gray line shows the measured sheet resistance for the 80 nm thick ITO layer. We measure a sheet resistance of 58.2 Ω/sq for the 80 nm thick layer of ITO sputtered on glass, in good agreement with literature values.[9,35] Figure 4 shows a decreasing sheet resistance for all networks as $w$ increases, ranging from 38.7 Ω/sq for the $\Lambda$ = 1000 nm, $w$ = 45 nm network down to 6.5 Ω/sq for the $\Lambda$ = 500 nm, $w$ = 110 nm network. The measured sheet resistance is well below that of ITO for all measured networks. The solid lines in Fig. 4 represent a fit according to $R_s = aw^{-1}$, where $a = \rho L / h$ for perfect Ohmic scaling. Good agreement is observed for the model, except for the networks with smallest $w$ where a higher resistance is found. Indeed, SEM images show higher defect densities and variations in wire width for the thinnest wires $w = 45$ nm. From the fitted values for $a$ the resistivity of the material $\rho$ is found for each pitch. Averaging the results gives $\rho = (9.3 \pm 0.4) \times 10^{-8}$ Ωm, which is ~5.7 times the bulk resistivity of silver ($1.59 \times 10^{-8}$ Ωm). The difference is attributed to structural wire defects and discontinuities in the wires, variations in the wire width, and electron scattering from grain boundaries.

**Discussion**

The optical and electrical measurements demonstrate that Ag nanowire networks are effective transparent conductors, for which optical transmittances and sheet resistances can be tuned by varying the wire diameter and pitch. Figure 5 summarizes the measured data, plotting the optical transmission averaged over the AM1.5 solar spectrum versus the sheet resistance for the three network pitches. The transmission and resistance data for the 80 nm ITO layer are also shown as horizontal and vertical gray dashed lines. From this graph three major trends can be observed. First, all NW networks show the inherent trade-off between optical transmittance and low sheet resistance: a lower sheet resistance can only be obtained at the expense of a reduced optical transmittance. Second, smaller-pitch networks have higher optical transmittance for a given sheet resistance. Similarly, for a transmittance that is equal to that of ITO, smaller-pitch networks have lower sheet resistance. Third, the smallest ($w = 45$ nm) $\Lambda = 500$ nm networks shows identical transmittance and lower sheet resistance than ITO; the $\Lambda = 700$ and $\Lambda = 1000$ nm networks show both higher transmittance and lower resistance than ITO.

Based on the four optical scattering processes described above, the advantage of using small $w$ is due to the fact that this: (1) shifts the LSPR wavelength to the blue where the AM1.5 solar spectrum is less intense; (2) reduces the coupling to the SPP and Rayleigh anomaly due to the smaller scattering cross section; and (3) shifts the MIM cut-off to longer wavelengths. The second loss mechanism; the excitation of NW SPPs is directly related to the NW network pitch. Larger pitch reduces the in-plane momentum obtained from the grating such that coupling to SPPs occurs at wavelengths outside the spectral range of interest. It also increases the clear aperture of the network and increases the MIM cut-off wavelength. However, the increase in optical transmission as a result of the ±1 diffraction orders directing into the plane for $\lambda > \Lambda$ and the linear relation between the sheet resistance and $\Lambda$ suggest the pitch should be small.

To gain insight in the influence of wire width and pitch on the overall performance as a transparent conductor, we show the figure of merit (FoM), defined as the ratio of the electrical conductance and the optical conductance ($\sigma_{DC}/\sigma_{opt}$), as a function of wire width, as an inset in Fig. 5. The definition of $\sigma_{DC}/\sigma_{opt}$ is given by[36,37]

$$T = \left(1 + \frac{188.5\,\sigma_{opt}}{R_S\ \sigma_{DC}}\right)^{-2}, \qquad (2)$$

where we have used the measured sheet resistance and averaged transmission for $R_S$ and $T$ respectively. The FoM of ITO is shown as a gray horizontal dashed line. This figure shows that the FoM of all NW networks is higher than that of ITO. Furthermore, the trends indicate that the optimum performance is expected for small pitch and small wire diameters.

For photovoltaic applications, the short minority carrier diffusion length in many materials, such as a-Si, suggests that a small pitch would be desirable for better carrier collection. In addition to this, the network pitch determines the amount of in-plane momentum ($m2\pi/\Lambda$) generated by the network, such that efficient light trapping can occur when the pitch is tuned to match the in-plane momentum of waveguide modes in the substrate. This is particularly relevant for thin-film solar cells. In summary, for application of transparent Ag nanowire networks in solar cells, the optimum design depends on the thickness of the substrate (which determines the waveguide modal dispersion), substrate refractive index, carrier diffusion length and absorption length of the solar cell.

Finally, we note that while the EBL technique used in this study provided an excellent tool to systematically study the metal nanowire networks, an inexpensive, large-area technique is needed for practical application. Recently, soft-nanoimprint techniques, such as substrate conformal imprint lithography (SCIL),[38,39] have become available to print metal nanostructures with the required dimensions over large areas in an inexpensive manner.

**Conclusion**

In conclusion, we have fabricated two-dimensional periodic silver nanowire networks with wire diameters in the range 45-110 nm and pitches of 500, 700 and 1000 nm. From a systematic study of optical transmission spectroscopy and numerical modeling we identify four physical mechanisms that determine the optical transmission of the network: (1) the LSPR on the individual nanowires; (2) diffractive coupling to the Rayleigh anomaly; (3) cut-off of the fundamental $TE_0$ mode in the MIM waveguides formed by the nanowires; and (4) coupling to SPPs on the wires parallel to the electric field component. Based on this, it is found that the best transparent nanowire network is obtained for thin wires and small pitch. We observe a normalized transmittance up to 91% averaged over the AM1.5 solar spectrum (larger than the clear aperture of network) for the best transmitting network and sheet resistances as low as 6.5 Ω/sq for the best conducting network. Our best performing networks show both better optical transparency and lower sheet resistances than an 80 nm thick ITO layer sputtered on glass. Finally, due to the two-dimensional periodic structure our networks can also function as a scattering layer, thereby combining the functionality of an electrical contact and a light trapping scheme in one design.

FIGURES

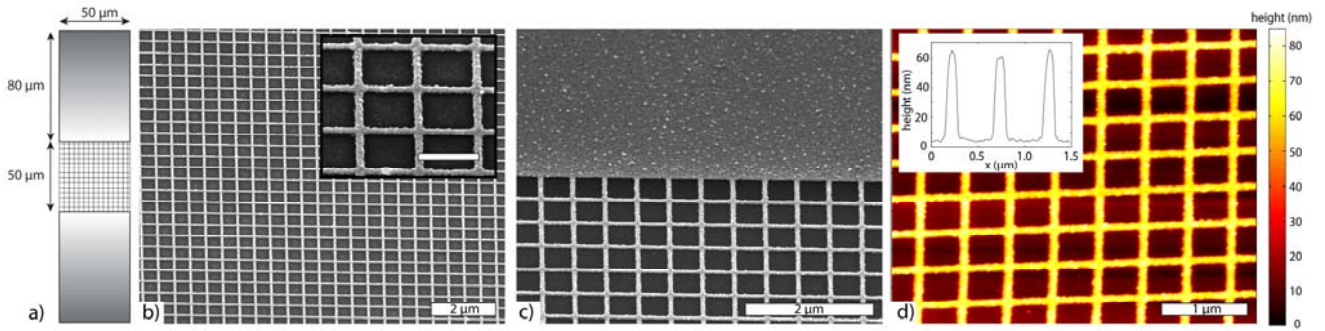

**Figure 1.** a) Sketch of the sample geometry. Silver nanowire networks, 50×50 μm in size, with wire diameters between 45-110 nm and a pitch of 500, 700 or 1000 nm were fabricated using electron beam lithography. Silver electrical contact pads with a size of 80×50 μm were fabricated on two opposing sides. The lines in the network are not to scale. b) SEM image of the $\Lambda$ = 500 nm, $w$ = 55 nm network showing the uniformity of the network over a large area. A large-magnification SEM image (inset) shows the high quality of the individual Ag wires. Scale bar in the inset is 500 nm. c) SEM image showing the contact area of the wire network and the electrical contact pads. d) Height profile (colorbar) of the $\Lambda$ = 500 nm, $w$ = 55 nm network obtained from AFM measurements. The inset shows a line scan from which the height of the wires is determined to be 60±5 nm. Note the different horizontal and vertical scales in the inset.

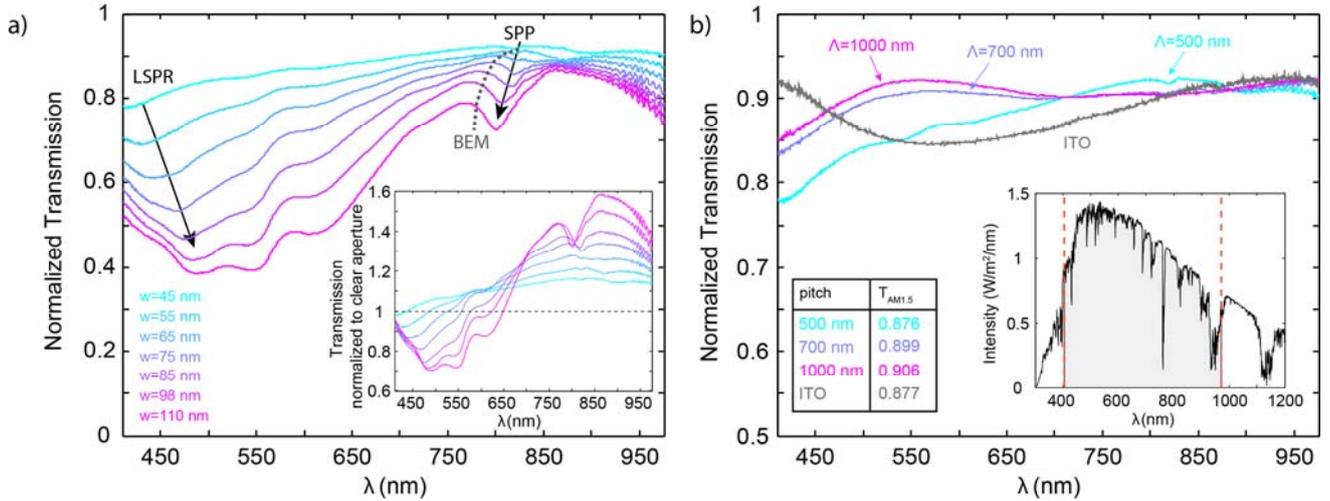

**Figure 2.** a) Normalized optical transmittance through nanowire networks with a pitch of 500 nm as a function of wavelength for different wire diameter *w* (colors). Arrows indicate the dips in transmission due to the excitation of LSPRs and SPPs. The inset shows the same transmission data normalized to the clear aperture. Anomalous transmission is observed for wavelengths above the LSPR wavelength. b) Normalized transmission as a function of wavelength for networks with the smallest wire diameter ($w$ = 45 nm) and pitch of 500 nm (blue), 700 nm (purple) and 1000 nm (pink). Also shown is the transmittance through an 80 nm thick layer of ITO sputtered on glass (gray). The inset shows the spectral intensity distribution of the AM1.5 solar spectrum, with the measured spectral region shaded in gray. The average transmission weighted for the AM1.5 spectrum is listed in the table as an inset.

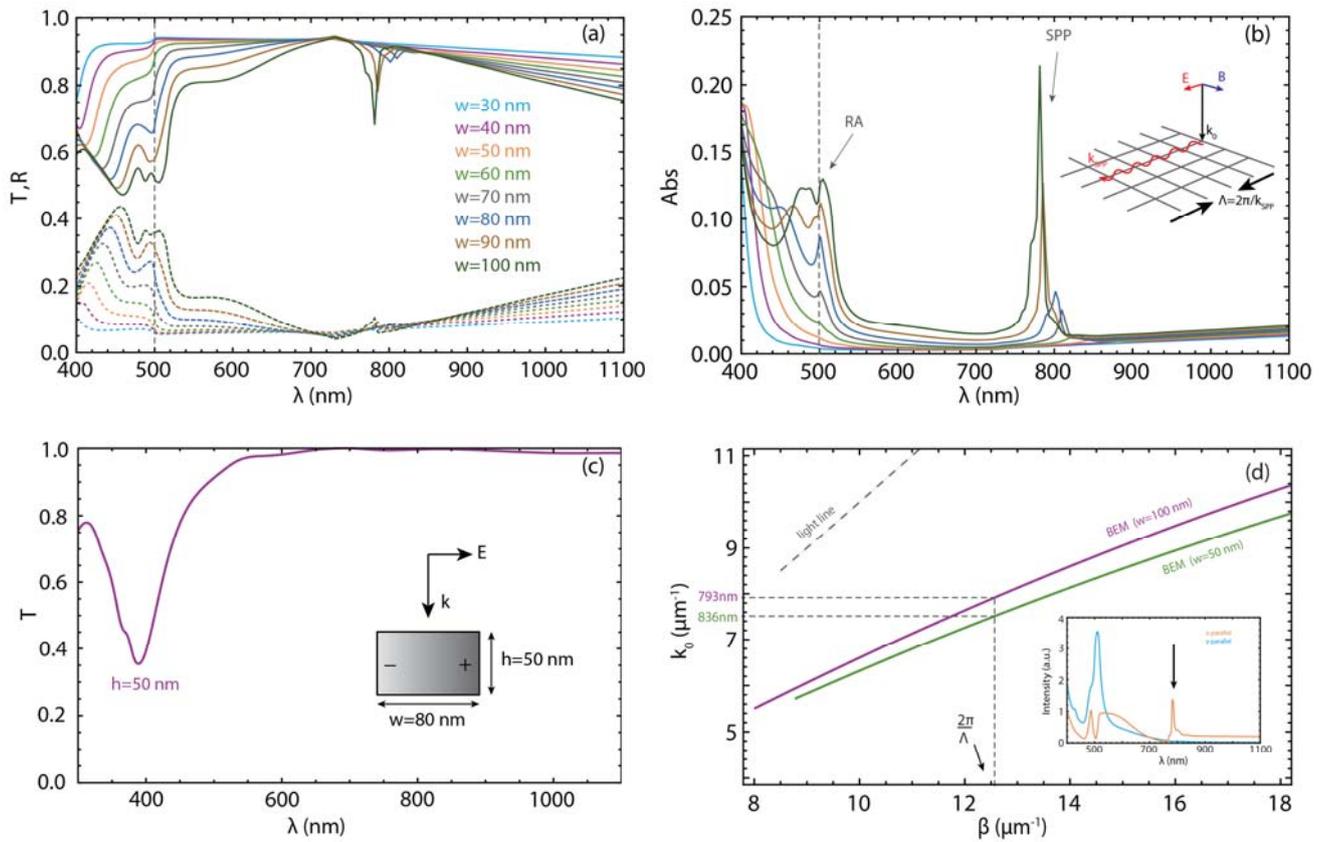

**Figure 3.** a) Simulated transmission (solid) and reflectance (dashed) as a function of wavelength for a range of wire widths (colors), for $h = 50$ nm. Also shown is the wavelength equal to the wire spacing (500 nm, vertical gray dashed line). b) Absorption in the nanowires, calculated as $Abs = 1-R-T$. Apart from the intrinsic absorption of silver in the blue, two peaks due to the Rayleigh anomaly (RA) and coupling to the surface plasmon polariton (SPP) are observed that are labeled with arrows. A sketch of the network is shown as an inset, with the SPP shown as a red oscillating arrow. c) Transmission as a function of wavelength for a single nanowire with $h = 50$ nm and $w = 80$ nm, located in air (simulated box width 400 nm). The inset shows a sketch of the cross section of the wire with a displaced free electron cloud (+ and – sign) as a result of the transverse free electron oscillation of the LSPR. d) Dispersion relation for a confined SPP mode on a wire with $w = 100$ nm (purple) and $w = 50$ nm (green) as calculated with BEM. The wavelength at which momentum matching occurs is shown (horizontal gray dashed lines). The inset shows the FDTD-calculated near-field intensity as a function of wavelength 5 nm above the wire parallel (orange) and perpendicular (blue) to the electric field component of the incoming light. The orange line shows a peak at the wavelength where light couples to the SPP mode, indicated by the arrow.

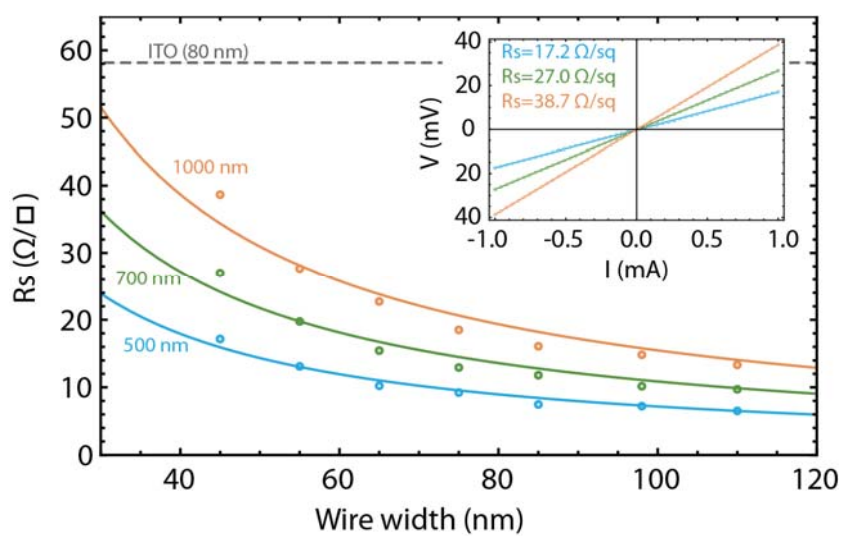

**Figure 4.** Measured sheet resistance (dots) as a function of wire width for different network pitch (500 nm (blue), 700 nm (green) and 1000 nm (orange)). The solid lines represent model fits for each pitch. Also shown is the measured sheet resistance of an 80 nm thick ITO layer sputtered on glass (gray dashed line). The inset shows measured I-V-curves for $w = 45$ nm for all three pitches (dots) with a linear fit (solid line).

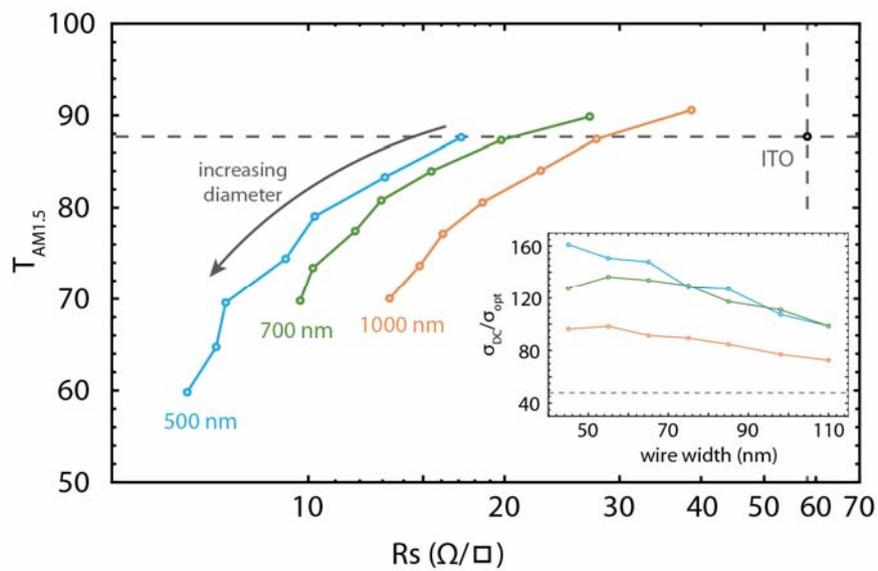

**Figure 5.** AM1.5 averaged transmission as function of sheet resistance for all networks with 500 nm pitch (blue), 700 nm pitch (green) and 1000 nm pitch (orange). The horizontal and vertical gray dashed lines show the transmittance and resistance of an 80 nm thick layer of ITO sputtered on glass. Note the logarithmic horizontal axis. The inset shows the Figure of Merit (FoM), defined as the ratio of the electrical conductance and the average optical conductance, as a function of wire width. The FoM of ITO is shown as a dashed gray horizontal line.

ACKNOWLEDGMENT: This work is part of the research program of the Foundation for Fundamental Research on Matter (FOM) which is financially supported by The Netherlands Organization for Fundamental Research (NWO). It is also supported by the Global Climate and Energy Project (GCEP) and the European Research Council.